\documentclass[12pt]{article}

    \def\+{{+\!\!\!+}}

    \def\d{\partial}
    \def\a{\alpha}
  \def\m{\mu}
  \def\n{\nu}
    \def\th{\theta}
    \def\g{\gamma}
    \def\G{\Gamma}
    \def\N{\nabla}
    
  \def\j{{\cal J}}
  \def\pcal{{\cal P}}
 \def\Lie{{\cal L}}
    \def\de{\delta}

    \def\r{\rho}
    \def\l{\lambda}
    \def\L{\Lambda}
    \def\s{\sigma}

    \def\e{\varepsilon}

    \def\implies{\Rightarrow}

          \def\b1{\hbox{I\hspace{-0.10in}1\hspace{0.04in}}}   %blackboard bold 1
    \def\pmb#1{\setbox0=\hbox{#1}%
    \kern.0em\copy0\kern-\wd0
    \kern-.04em\copy0\kern-\wd0
    \kern.08em\copy0\kern-\wd0
    \kern-.04em\raise.0433em\box0 }         %poor man's bold macro (TexBook)
    \def\half{\frac{1}{2}}

    \newcommand{\nc}{\newcommand}
    \nc{\beq}{\begin{equation}}
    \nc{\eeq}[1]{\label{#1}\end{equation}}
    \nc{\ber}{\begin{eqnarray}}
    \nc{\eer}[1]{\label{#1}\end{eqnarray}}
    \nc{\pek}[1]{\cite{#1}}
    \nc{\enr}[1]{(\ref{#1})}
    \nc{\kal}[1]{{\cal{#1}}}
    \nc{\dott}{\;\cdot\;}
    \newcommand{\be}{\begin{equation}}
    \newcommand{\ee}{\end{equation}}
    \newcommand{\bea}{\begin{eqnarray}}
    \newcommand{\eea}{\end{eqnarray}}
    \newcommand{\re}[1]{(\ref{#1})}

\begin{document}

\title{Comments on Dualities on Group Manifolds and Supersymmetry}

\author{ Svend E. Hjelmeland\thanks{s.e.hjelmeland@fys.uio.no}\,\,$^a$ and
 Ulf Lindstr\"om\thanks{ulf.lindstrom@teorfys.uu.se}\,\,$^{b}$
\\
\, \\
{\small{$^a$Department of Physics, University of Oslo,}}\\
{\small{P.O.Box 1048 Blindern, N-0316 Oslo, Norway}}\\
\, \\
{\small{$^{b}$ITP, University of Uppsala,}}\\
{\small{ Box 803, S-75108  Uppsala, Sweden}}\\}

%\date{\today}

\maketitle

\begin{picture}(0,0)(0,0)
\put(300,370){OSLO-TP 6-03} \put(300,355){ITP-UU-10-03}
\put(300,340){hep-th/xxxyymm}
\end{picture}
\vspace{-24pt}

\begin{abstract}
We review some facts about various T-dualities and sigma models on
group manifolds, with particular emphasis on supersymmetry. We
point out some of the problems in reconciling Poisson-Lie duality
with the bi-hermitean geometry of N=2 supersymmetric sigma models.
A couple of examples of supersymmetric models admitting
Poisson-Lie duality are included.
\end{abstract}

\newpage
\tableofcontents
\section{Introduction}

Supersymmetric sigma models are of interest, e.g.,  as gauge-fixed
string actions, as representing exact string vacuua (WZW-models),
for their intimate connection to complex geometry (of the target
manifold) and for their role as effective low-energy actions for
supergravity scalars.

The various(generalized) T-dualities for sigma models are
important in the context of strings, where, e.g., the usual
T-duality relates different geometries describing one and the same
physical configuration.

In this article, we review some facts about various T-dualities
with emphasis on sigma models on group manifolds. Particular
emphasis is put on the requirements for $N=(2,2)$ supersymmetry,
(bi-hermitean target space). We comment on the situation for
general models not describable as WZW models. The latter half of
the paper consists of a discussion of Poisson-Lie duality for
$N=1$ supersymmetric nonlinear sigma models and examples based on
$SU(2)\otimes U(1)$.

This paper grew out of an effort to understand how the stringent
requirements on the target space geometry of $N=(2,2)$
supersymmetric sigma models might be made to agree with
Poisson-Lie duality in a more general case than abelian or
non-abelian T-duality. While we have not resolved the original
problem, we believe that the discussion contained in this paper
will serve as a necessary background and starting point. Along the
way we have collected a number of observations and comments which
may find their use in other contexts as well.

 \section{Nonlinear Sigma Models}

  In this section we collect the necessary background on
  (supersymmetric) two-dimensional non-linear sigma models.

  The action for a general non-linear sigma model is
  \beq
  S=\int d^{2}\xi
  \partial X^{\m}E_{\m\n}(X)
  \bar{\partial}X^{\n}
  \eeq{boseaction}
  where the metric $G_{\m\n}=\frac{1}{2}E_{(\m\n)}$ and the
  torsion potential
  $B_{\m\n}=\frac{1}{2}E_{[\m\n]}$ \footnote{ We use (anti-) symmetrization
 without a combinatorial factor}.

  In $N=1$ superspace this becomes
 \begin{equation}
  \label{susyaction}
  {\cal S}=i\int d^{2}\xi d^{2}\theta
  D_{+}X^{\m}E_{\m\n}D_{-}X^{\n}
  \end{equation}
  where again $E_{\m\n}=G_{\m\n}+B_{\m\n}$ and where $G_{\m\n}(X)$ and $B_{\m\n}(X)$  and
  $X^\m(\xi,\th)$ are superfields
  whose lowest components enters in (\ref{boseaction}) above, (we
  use the same notation for superfields as for their lowest
  components).

  As first described in \cite{Gates:nk},
  the action (\ref{susyaction}) has $N=(2,2)$ supersymmetry
  \footnote{The target-space geometry for models with less supersymmetry,
  e.g, $(2,1)$, is also very interesting, but will not be discussed here. See \cite{Abou-Zeid:1999em}.} i.e.,
  an additional non-manifest supersymmetry of the form
  \beq
  \de X^\m =
  \e^+(D_+X^\n){\j}_\n^{(+)\m}+\e^-(D_-X^\n){\j}_\n^{(-)\m}~,
  \eeq{secondSUSY}
  provided that $\j^{(\pm)}$ are complex structures:  they square to minus
  one,
  \beq
  \j^{2(\pm)}=-{\b1}~,
  \eeq{Cstruc}
  and have vanishing Nijenhuis tensors\footnote{More general models with non-vanishing ${\cal N}$ have
  also been considered \cite{Delius:nc}};
  \beq
  {\cal N}_{\m\n}^{~(\pm)\kappa}\equiv\j_\m^{(\pm)\g}\d_{[\g}\j_{\n]}^{(\pm)\kappa}-(\m\leftrightarrow
  \n)=0~.
  \eeq{Nij}
  In addition, the metric has to be {\em bi-hermitean}, i.e., hermitean with respect to {\em both} complex
  structures
  \beq
  \j_\m^{(\pm)\g}G_{\g\r}\j_\n^{(\pm)\r}=G_{\m\n}~,
  \eeq{biherm}
  and the complex structures should be covariantly constant with respect to certain
  connections $\G^{(\pm)}$, respectively
  \beq
  \N_\m^{(\pm)}\j_\n^{(\pm)\g}=0~.
  \eeq{CovJ}
  These  connections are
  \beq
  \G_{\m\n}^{(\pm)\g}=\G_{\m\n}^{(0)\g}\pm T_{\m\n}^{~\g}~,
  \eeq{Connex}
  with $\G^{(0)}$ the Christoffel connection for the  metric $G$, and the
 torsion
  given by
  \beq
  T_{\m\n}^{~\g}=H_{\m\n\r}G^{\r\g}~.
  \eeq{tor}
  This  relates the complex structures to the field-strength for the $B$-field,
  \beq
  H_{\m\n\r}=\d_{[\m}B_{\n\r]}~,
  \eeq{H}
  which implies
 \beq
H_{\m\n\r}=\j_\m^{(+)\g}\j_\n^{(+)\kappa}\j_\r^{(+)\l}d\j^{(+)}_{\g\kappa\l}=-\j_\m^{(-)\g}\j_\n^{(-)\kappa}\j_\r^{(-)\l}d\j^{(-)}_{\g\kappa\l}~,
  \eeq{HJrel}
  where $d\j^{(\pm)}$ is the exterior derivative of the two forms with
 components
  $\j_{\m\n}^{(\pm)}=\j_\m^{(\pm)\g}G_{\g\n}$, (antisymmetrical because of
  (\ref{biherm}))\footnote{For a recent discussion of the relevant geometry, see \cite{Lyakhovich:2002kc}}.

  In fact, the set of conditions (\ref{Nij})-(\ref{H}), derived from
 requiring
  that the
  action (\ref{susyaction}) is invariant under the variation (\ref{secondSUSY})
  along with
  closure of the algebra, is a minimal set of requirements and may be modified
 as
  seen from
  the following.

  Condition (\ref{HJrel}) may be equivalently expressed as a relation which
  states
  that the complex structures preserve the torsion, i.e., the field strength of
  the
  $B$-field. We may strengthen this condition to preservation of the $B$-field
  itself by
  replacing condition (\ref{biherm}) by
  \beq
  \j_\m^{(\pm)\g}E_{\g\r}\j_\n^{(\pm)\r}=E_{\m\n}~,
  \eeq{Ecndn}
  since the antisymmetric part of this reads
  \beq
  \j_\m^{(\pm)\g}B_{\g\r}\j_\n^{(\pm)\r}=B_{\m\n}~.
  \eeq{Bcndn}
  (The symmetric part is (\ref{biherm}).) Imposing (\ref{Ecndn}) in the
 variation
  of
  (\ref{susyaction}), we find that (\ref{CovJ}) is weakened to
  \ber
 && E_{[\m|\n}\d_\r\j^{(+)\n}_{|\l]}-\d_{[\m|}E_{\n\r}\j_{|\l]}^{(+)\n}+\d_\n
  E_{[\m|\r}\j_{|\l]}^{(+)\n}=0\\
&&  E_{\n[\m|}\d_\r\j^{(-)\n}_{|\l]}-\d_{[\m|}E_{\r\n}\j_{|\l]}^{(-)\n}+\d_\n
  E_{\r[\m|}\j_{|\l]}^{(-)\n}=0~.
  \eer{NJcond}
  We recover the previous conditions as follows: Combining (\ref{NJcond}) with
  the
  derivative of (\ref{Ecndn}) returns (\ref{CovJ}). Antisymmetrizing
  (\ref{NJcond}) in all
  three indices and multiplying with
  $\j_\m^{(\pm)\g}\j_\n^{(\pm)\kappa}\j_\r^{(\pm)\l}$
  yields (\ref{HJrel}). Hence the new conditions represent a special case of the
  general
  structure.

  Since we have a condition (\ref{Ecndn}) which is stronger than
  necessary for
  $N=2$, we may ask if it is compatible with other conditions on the theory.
We first note that when the two complex structures commute,
$[\j^{(+)},\j^{(-)}]=0$, their product gives an almost product
structure, i. e., $\pmb{$\Pi$}^2=\b1$ where
\pmb{$\Pi$}$=\j^{(+)}\j^{(-)}$ \cite{Gates:nk}. While the
individual integrability of $\j^{(+)}$ and $\j^{(-)}$ is not
sufficient to guarantee integrability of \pmb{$\Pi$}, in
conjunction with \re{CovJ} it is \cite{Gauntlett:2003cy}. We may
then choose coordinates where \bea \pmb{$\Pi$}=
\left(\begin{array}{cc} {\b1} & 0\\ 0 & -{\b1}
\end{array}\right)~.\eea
\label{prods} It is shown in \cite{Gates:nk} that the subspaces
projected out by $\hat P_{\pm}\equiv \half(\b1\pm\pmb{$\Pi$})$ are
K\"ahler, i.e., these sectors contain no $B$-field. Since
\re{Ecndn} implies $\hat P_+E\hat P_-^T=0$ it also follows that
$E$ has no ``mixed'' components in these coordinates. Hence, we
conclude\footnote{Using the explicit forms in \cite{Gates:nk} one
verifies that $\j^{(-)}$ indeed does not preserve $B$.} that
\re{Ecndn} is compatible with $B\ne 0$ only if
$[\j^{(+)},\j^{(-)}]\ne 0$. This excludes formulations in terms of
chiral and twisted chiral superfields \cite{Gates:nk}, but may
allow (anti-)semichiral superfields \cite{Buscher:uw} as
coordinates, as discussed in \cite{Ivanov:ec}. (For $N=4$ the
geometric structure is even more restricted \cite{Gates:nk}, and
there are additional superfield coordinates available
\cite{Lindstrom:mw}.)

 Another property one might want to study is the gauge transformation of
 the $B$-field. In the absence of
  boundaries, at
  least, this field only enters the field equations through its field strength
  (\ref{H}).
  Under what conditions is that compatible with (\ref{Bcndn})? To answer this we
  define the
  projection operators
  \beq
  \pcal_{\pm}^{(\pm)}\equiv{\frac 1 2}\left(\b1\pm i\j^{(\pm)}\right)~.
  \eeq{Proj}
  Using these, we may restate \re{Ecndn}, or equivalently (\ref{biherm}) and (\ref{Bcndn}), as expressing
  \beq
  \pcal_{\pm \m}^{~~\n}\pcal_{\pm \s}^{~~\r}E_{\n\r}=0, \implies
  \pcal_{\pm \m}^{~~\n}\pcal_{\pm \s}^{~~\r}G_{\n\r}=
  \pcal_{\pm \m}^{~~\n}\pcal_{\pm \s}^{~~\r}B_{\n\r}=0~.
  \eeq{CanGcomp}
  (The $(\pm)$-labels distinguishing the two complex structures
 are
  supressed
  for easier readability). These relations mean that in the canonical
 coordinates
  of the
  complex structures the tensors have only mixed components, $G_{i\bar
  j}$,
  $G_{{\bar i}j}$,
  $B_{i\bar j}$ and $B_{\bar i j}$. A gauge transformation of $B$ infinitesimally
  reads $\de
  B_{\m\n}=\d_{[\m}\L_{\n]}$. We deduce that we must have
  \beq
  \pcal_{\pm \m}^{~~\n}\pcal_{\pm \s}^{~~\r}\d_{[\n}\L_{\r]}=0~,
  \eeq{Cangauge}
  or $\d_{[i}\L_{ j]}=\d_{[\bar i}\L_{\bar j]}=0$. A strong version of this condition is that the
  transformation
  parameters  have to have (anti-) holomorphic components $\L_i (\bar z)$,
  $(\L_{\bar i}( z)$) in both sets of canonical coordinates. (Of
  course $\L_i =\d_i\L$ and c.c. also solve the constraints.)

This shows that the requirement of $N=2$ supersymmetry leads to restrictions on the gauge
symmetry for the $B_{\m\n}$-field. We shall see in Sec. 5 below, that this symmetry is also in
conflict with the Poisson-Lie condition.

 \section{Sigma Models on Group Manifolds}

We shall be particularly interested in non-linear sigma models on
group manifolds. For open models, interesting relations between
the geometry and boundary conditions were discovered
in\footnote{Based on previous results for general supersymmetric
sigma models in \cite{Albertsson:2001dv,Albertsson:2002qc} }
\cite{Lindstrom:2002jb,Lindstrom:2002vp,Albertsson:2003va} and,
generally, they are the appropriate setting for Poisson-Lie
duality, which we discuss in Sec. 5.
 In a group $G$, we parametrize the group elements $g\in G$ using coordinates
 $X^\mu$, and define the left and right frames by
 \beq
 g^{-1} \de g=L_\mu\de X^\mu ;\qquad \de gg^{-1}=R_\mu\de X^\mu~,
 \eeq{LeRi}
 where $L_\mu\equiv L_\mu^AT_A\quad R_\mu\equiv R_\mu^AT_A$, with $T_A$ the
 generators of the corresponding Lie algebra
 $[T_A,T_B]=f_{AB}^{~~C}T_C$. In these coordinates, a general sigma model on the
 group space may be written
 \beq
  {\cal S}=i\int d^{2}\xi d^{2}\theta
  D_{+}X^{\m}E_{\m\n}D_{-}X^{\n}=i\int d^{2}\xi d^{2}\theta Tr (g^{-1}
 D_{+}g)^AE_{AB}(g^{-1} D_{-}g)^B
  \eeq{Gaction}
  where $E_{\m\n}= L_\m^AE_{AB}L^B_\n$. In the special case of a
 Wess-Zumino-Witten model we also have (in the bosonic sector)
 \beq
  {\cal S}=\int_{\d Y} d^{2}\xi Tr (g^{-1} \d g)(g^{-1} \bar\d g)
 +{\frac 1 3}\int_{Y}  Tr (g^{-1} dg\wedge g^{-1} dg\wedge g^{-1} dg )~,
  \eeq{WZW}
 which means that the symmetric part of $E_{AB}$ is the Killing form and that
 the
 torsion is $H_{\m\n\r}=L_\m^AL_\n^BL_\r^Cf_{ABC}$, where $f_{ABC}$ are the
 completely antisymmetric structure constants. Further, the two-dimensional
 space
 $\d Y$ has been extended to $Y$ with an additional auxiliary coordinate. For
 this case one can show that the $N=2$ conditions in Sec 1. above require
 \beq
 \N_\r^{(-)}L_A^\m=0~, \qquad \N_\r^{(+)}R_A^\m=0~.
 \eeq{NabLR}
 In the more general case we want to consider here, we derive the following
 relation for the covariant derivatives of the left frames:
 \ber
 &&\N^\r L_A^\m=-{\frac 1 2}\left(
 f_A^{~(\r\m)}+f^{\r\m}_{~~~A}+2T_A^{~(\r\m)}+2T^{\r\m}_{~~~A}+L_B^\r L_C^\m\N
 _AG^{BC}+L^{B[\r}\N^{\m]}G_{AB}\right)\\
&& \N^\r R_A^\m=-{\frac 1 2}\left(
 -f_A^{~(\r\m)}-f^{\r\m}_{~~~A}+2T_A^{~(\r\m)}+2T^{\r\m}_{~~~A}+R_B^\r R_C^\m\N
 _AG^{BC}+R^{B[\r}\N^{\m]}G_{AB}\right)~,\nonumber\\
 \eer{NabLRgen}
 where Lie-algebra indices are raised and lowered with $G_{AB}={\frac 1
 2}E_{(AB)}$, transformed into target space indices with $L_A^\m$ or  $R_A^\m$, and we have
 allowed for a torsion $T$ and assumed that $\N_\m G_{\n\r}=0$. Target space indices are raised and lowered
 using $G_{\m\n}$.
 Clearly, when $G_{AB}$ is constant and the structure constants $f_A^{~(\r\m)}+f^{\r\m}_{~~~A}$
 equal to ($\mp 2$
 times) the torsion, we recover (\ref{NabLR}). (In the WZW case, in addition,
$f_{ABC}\equiv f_{AB}^{~~~D}G_{DC}$ are completely antisymmetric.)

 \section{Isometry-based T-Dualities}

The idea of dual formulations describing the same physical situation had been around a long
time in the context of sigma models when it found its application in string theory. In fact the
geometry changing aspects make it particularly interesting for two-dimensional models, but there
are many features that are fascinating in general. See, e.g.,
\cite{Giveon:1994mw}-\cite{Hassan:1995je} for reviews and general aspects of sigma model duality.

 When the sigma model
\re{boseaction} or \re{susyaction} has (generalized)
 isometries\footnote{The conditions stated here are stronger than necessary. Typically we need
only require $\Lie_{ k}B=d\omega$. For a thorough discussion of
conditions on isometries and their gauging in the context of susy
WZW models, see \cite{Hull:ms}. } with Killing vector fields
$k_A=k_A^\m\d_\m$ with algebra $[k_A,k_B]=f_{AB}^{~~C}k_C$
 \beq
 \de X^\m=\e^Ak_A^\m=\Lie_{\e k}X^\m~,\qquad\Lie_{\e k}E_{\m\n}=0~,
 \eeq{isom}
 there exists a ``parent action'' from which the sigma model and its T-dual can
 be derived. In the bosonic sector it reads
 \beq
  {\cal S}=\int d^{2}\xi \left(DX^{\m}E_{\m\n}(X)
  \bar{D}X^{\n}+tr\L F\right)~,
 \eeq{Parent}
 where the covariant derivatives and the field strength are
 \beq
 DX^\m=\d X^\m+A^Ak_A^\m~, \quad \bar{D}X^\m=\bar{\d}X^\m+\bar{A}^Ak^\m_A~, \qquad
 F=[D,\bar{D}]=\d\bar{A}-\bar{\d}A+[A,\bar{A}]~.
 \eeq{covF}
 Varying $\L$ gives that $F$ is pure gauge. Plugging this back into \re{Parent}
 we recover the original action \re{boseaction}, whereas the $A$-field equations
 instead yield the dual action in terms of $\L$. For abelian isometries this
 prescription is unproblematic. For one isometry and in coordinates adapted to
 this isometry it yields the famous Buscher rules \cite{Buscher:sk}\cite{Buscher:qj} relating the original
 background
 $G,B$ to the dual background $\tilde G, \tilde B$:
 \ber \nonumber
 &&\tilde G_{00}=G_{00}^{-1}~,\qquad \tilde G_{0i}=G_{00}^{-1}B_{0i}~,\\
 \nonumber
 &&\tilde G_{ik}=
 G_{ik}-G_{00}^{-1}\left(G_{i0}G_{0k}+B_{i0}B_{0k}\right)~,\\
 \nonumber
 &&\tilde B_{ik}=
 B_{ik}+G_{00}^{-1}\left(G_{i0}B_{0k}+B_{i0}G_{0k}\right)~,\\
 &&\tilde B_{0i}=G_{00}^{-1}G_{0i}~,
 \eer{Buscher}
 where $0$ is the isometry direction and $i$ denotes the rest of the coordinates
 (the spectators). With the appropriate superfield interpretation, these rules apply also to $N=1$ supersymmetric models.

The relations \re{Buscher} are expressed in adapted coordinates where $G$ and $B$ are independent
of the isometry direction $X^0$, (although one may formulate the rules in a covariant fashion
using the Killing vectors). It is interesting that in the dualization for $N=2,4$ models in
superspace, which is achieved via a gauging of holomorphic isometries \cite{Hull:1985pq}, the
dual model is described directly in canonical complex coordinates \cite{Lindstrom:rt}. This is
related to the fact that there duality relates the K\"ahler potentials rather than the metric.

The above rules also generalize to the case of several commuting isometries.

Several items may be mentioned at this stage. Firstly, as is
obvious from the factors of $G_{00}^{-1}$, the case of a lightlike
isometry has to be treated separately. Secondly, although
T-duality is always compatible with supersymmetry, it is sometimes
necessary to take non-local world-sheet effects into account
\cite{Bakas:1996zp},\cite{Hassan:1995pk}. Thirdly, typically the
complex geometries of the $N=2,4$ target spaces will only be
preserved if the isometries active in the duality commute with the
supersymmetries.

Non-abelian duality generalizes the above relations for the case of a non-abelian isometries
\cite{delaOssa:1992vc}. It is somewhat more problematic, partly due to the fact that
 the dual of the dual model does not return the original, i.e., unlike the
 abelian case the non-abelian duality is not idempotent. It is perhaps best
 studied within the framework of Poisson-Lie duality, which we now describe.

\section{Poisson-Lie Duality}

A very interesting generalization of T-duality to the situation when the group action is not an
isometry of the sigma model was constructed in \cite{kn:KS}, and has since been discussed
extensively, e.g., in
\cite{Klimcik:1995kw},\cite{Klimcik:1995dy},\cite{Alekseev:1995ym},\cite{kn:TvU},\cite{VonUnge:2002ss},
\cite{Sfetsos:1997pi},\cite{Sfetsos:1998kr}. Supersymmetric versions are treated in, e.g.,
\cite{Klimcik:1997vp}, \cite{Parkhomenko:1996ff},\cite{Sfetsos:1996xj},
\cite{Hjelmeland:2001ar}.

\subsection{Definitions}
 In Poisson-Lie duality the isometries in \re{isom} are generalized to the
 following relation
 \beq
 \Lie_{A}E_{\m\n}\equiv\Lie_{R_A}E_{\m\n}=-E_{\m\r}R^{\r}_{B}\tilde{f}^{BC}_{A}
 R^{\s}_{C}E_{\s\n}~.
 \eeq{PLcnd}

  Here $\tilde{f}^{AB}_{C}$ are structure constants in a dual Lie algebra.
For the sake of greater clarity we will not consider spectators, i.e., we will
 only keep the target space coordinates affected by the transformations
 \re{PLcnd}. We are thus effectively studying a $\s$-model on the corresponding group space.

 Klim\v{c}ik and \v{S}evera \cite{kn:KS}, show that the condition
 \re{PLcnd} can be solved
 and the dual model found provided that the Lie algebra ${\cal G}$ and its dual
 $\tilde{\cal G}$ form what is called a Drinfel'd double
 \cite{kn:D,kn:FG,kn:AM}.

 Let $G$ and $\tilde{G}$ be  groups obeying \re{PLcnd} on the original $\sigma$-model
 and its dual, respectively, with $\mbox{dim}G=\mbox{dim}\tilde{G}$.
 The corresponding Lie algebras are ${\cal G}$ and $\tilde{\cal G}$.
 Then the Drinfel'd double ${\cal D}^2\equiv G\otimes\tilde{G}$ and
 comes equipped with an invariant inner product $\langle~,~\rangle$.
 The corresponding algebra ${\pmb d}$
 consists of the two subalgebras
 ${\cal G}$ and $\tilde{\cal G}$ that are null-spaces w.r.t. this product.
 We choose two sets of generators $\{T_{A}\}$ and $\{\tilde T^{A}\}$ so that
 $\{T_{A}\}$ span ${\cal G}$ and $\{\tilde T^{A}\}$ span $\tilde{\cal G}$.
 The set $T_{{\cal A}}\in\{T_{A},\tilde T^{B}\}$ then span ${\pmb d}$. The Lie algebra of
 the Drinfel'd double generated by $T_{A}$ and $\tilde T^{A}$
 $(A=1,\ldots,\mbox{dimG})$, is
 \begin{eqnarray}
 \label{algebra}
 [T_{A},T_{B}]&=&f_{AB}^{C}T_{C}, \nonumber \\
 ~[\tilde T^{A},\tilde T^{B}]&=&\tilde{f}^{AB}_{C}\tilde T^{C}, \nonumber \\
 ~[T_{A},\tilde T^{B}]&=&\tilde{f}^{BC}_{A}T_{C}-f_{AC}^{B}\tilde T^{C},
 \end{eqnarray}
 where $f_{AB}^{C}$ and $\tilde{f}^{AB}_{C}$ are the structure constants of
 ${\cal G}$ and $\tilde{\cal G}$, respectively, and satisfy
 the Lie bi-algebra $({\cal G},\tilde{\cal G})$ consistency condition
 \begin{equation}
 \label{consistency}
 f^{A}_{DC}\tilde{f}^{RS}_{A}=
 \tilde{f}^{AS}_{C}f^{R}_{DA}+\tilde{f}^{RA}_{C}f^{S}_{DA}
 -\tilde{f}^{AS}_{D}f^{R}_{CA}-\tilde{f}^{RA}_{D}f^{S}_{CA}.
 \end{equation}
 This condition arises in the PL duality context as the condition
 $[{\cal L}_{k_{A}},{\cal L}_{k_{B}}]=f_{AB}^{~~C}{\cal L}_{k_{C}}$ applied
 to \re{PLcnd}.
 The invariant inner product between the generators has the following properties
 \begin{equation}
 \label{innerprod}
 \langle T_{A},T_{B}\rangle =\langle \tilde T^{A},\tilde T^{B}\rangle =0,
 ~~~~~~\langle T_{A},\tilde T^{B}\rangle =\delta_{A}^{~B}
 \end{equation}
 and obeys the invariance condition
 \begin{equation}
 \langle {\cal X}T_{{\cal A}}{\cal X}^{-1}, T_{{\cal B}}\rangle =
 \langle T_{{\cal A}},{\cal X}^{-1}T_{{\cal B}}{\cal X}\rangle,
 \end{equation}
 where ${\cal X}$ is any element of the Drinfel'd double or one of its
 subgroups.

 We define,
 \begin{eqnarray}
 \label{def}
 \mu^{AB}(g)&=&\langle g\tilde T^{A}g^{-1},\tilde T^{B}\rangle ;\ \ \ \ \
 \nu^{A}_{~B}(g)=\langle g\tilde T^{A}g^{-1},T_{B}\rangle; \nonumber \\
 \alpha_{B}^{~A}(\tilde{g})
 &=&\langle \tilde{g}T_{B}\tilde{g}^{-1},\tilde T^{A}\rangle ;\ \ \ \ \
 \beta_{AB}(\tilde{g})=\langle \tilde{g}T_{A}\tilde{g}^{-1},T_{B}\rangle
 \end{eqnarray}\label{mnab}
 which obey $\mu(g^{-1})=\mu^{t}(g)$, $\nu(g^{-1})=\nu^{-1}(g)$,
 $\alpha(\tilde{g}^{-1})=\alpha^{-1}(\tilde{g})$
 and $\beta(\tilde{g}^{-1})=\beta^{t}(\tilde{g})$ where $t$ stands for
 transpose.

 We return to the solution of (\ref{PLcnd}) given by Klim\v{c}ik and
 \v{S}evera. With $E_{\m\n}=L^{A}_{\m}E_{AB}L^{B}_{\n}$ as
 in (\ref{Gaction}) the solution is
 \beq
 E_{AB}=((E^{0})^{-1}+\Pi)^{-1}_{AB};\ \ \ \ \
 \Pi^{AB}=\mu^{AC}\nu_{C}^{~B}~,
 \eeq{KSsln1}
where $E^0$ is a constant matrix.
 Similarly, in the dual theory one has relations corresponding to
 (\ref{PLcnd}) and (\ref{Gaction}) and
 \beq
 \tilde{E}^{AB}=[(E^{0}+\tilde{\Pi})^{-1}]^{AB};
 \ \ \ \ \ \ \tilde{\Pi}_{AB}=\beta_{AC}\alpha^{C}_{~B}.
 \eeq{KSsln2}
 The abelian and non-abelian dualities described previously are
 special cases of the more general PL duality. In the non-abelian
 case we have $\mu^{AB}=0$, $\alpha^{A}_{~B}=\delta^{A}_{~B}$ and
 $\beta_{AB}=f^{C}_{AB}\tilde{x}_{C}$, where $\tilde{x}_{C}$ is the dual
 non-inert coordinates, so that $E_{AB}=E^{0}_{AB}$ and
 $\tilde{E}^{AB}=[(E^{0}+f^{C}\tilde{x}_{C})^{-1}]^{AB}$.

We now include spectators and give the generalized Buscher rules
(in the notation of \cite{Hjelmeland:2001ar})
\begin{eqnarray}
\label{genBT}
\tilde{E}^{-1}-\beta\alpha=(E^{-1}+\mu\nu)^{-1}&=&E^{0}(x^{\alpha});
\nonumber \\ 
\tilde{E}^{-1}\tilde{\cal F}^{R}=E^{0}E^{-1}{\cal F}^{R}&=&F^{R}(x^\a); \nonumber \\ 
-\tilde{{\cal F}}^{L}\tilde{E}^{-1}={\cal F}^{L}E^{-1}E^{0}&=&F^{L}(x^\a); \nonumber
\\ \tilde{F}-\tilde{\cal F}^{L}\tilde{E}^{-1}\tilde{\cal F}^{R}=
F+{\cal F}^{L}(E^{-1}E^{0}E^{-1}-E^{-1}){\cal F}^{R}&=&\hat{F}(x^\a)~.
\end{eqnarray}
Here the indices are split according to $\m\to(\hat i, \a)$, with
$\a$ representing the spectators. To be able to use a condensed
notation, we have replace $E$ by $F$ when it carries curved
indices. Further, ${\cal F}_{\alpha B}^{L}\equiv
F_{\alpha\hat{j}}L^{\hat{j}}_{~B}$, ${\cal F}^{R}_{A\beta}\equiv
L_{A}^{~\hat{i}}F_{\hat{i}\beta}$, $\tilde{{\cal
F}}_{\alpha}^{LB}\equiv
\tilde{F}_{\alpha}^{~\hat{j}}\tilde{L}_{\hat{j}}^{~B}$ and
$\tilde{{\cal F}}^{RA}_{~~\beta}\equiv
\tilde{L}^{A}_{~\hat{i}}\tilde{F}^{\hat{i}}_{~\beta}$.

These relations apply verbatim also to $N=1$ models \cite{Hjelmeland:2001ar}. For $N=2$, the general
rules that take into consideration the bi-hermitean geometry have not been worked out. In fact,
the whole Poisson-Lie structure is easily applied to $N=1$ models, whereas for  $N=2$ only
certain superconformal models have lent themselves to a Poisson-Lie description
\cite{Parkhomenko:1996ff}.

\subsection{The Poisson-Lie Condition Rewritten}

The Poisson-Lie condition \re{PLcnd} can be rewritten in a form
from which its solution may be found via integration. In a
particular case this may turn out to be just as efficient as
calculating the objects that enter the general solution
\re{KSsln1} and \re{KSsln2} above. Using the definition of the Lie
derivative of the frame fields and the fact that the left and
right fields commute, $[R_{A},L_{B}]^{\m}=0$, we find \beq {\cal
L}_{A}E_{\m\n}=R^{\r}_{A}L^{B}_{\m}(\d_\r E_{BC})L^{C}_{\n}
\eeq{LiderE} It follows that the Poisson-Lie condition can be
rewritten as\footnote{Note that $LR$ is the group element $ g$ in
the adjoint representation.}
\begin{equation}
R^{\r}_{A}\partial_{\r}E_{BC}=
-E_{BD}L^{D}_{\r}R^{\r}_{E}\tilde{f}^{EF}_{A}R^{\l}_{F}L^{G}_{\l}E_{GC}
\end{equation}
or in terms of the inverse matrix elements
\beq
R^{\r}_{A}\d_{\r}E^{DE}=
L^{D}_{\r}R^{\r}_{B}\tilde{f}^{BC}_{A}R^{\l}_{C}L^{E}_{\l}
\eeq{inversePL}
The dual relation is
\begin{equation}
\tilde{R}_{\r}^{A}\partial^{\r}\tilde{E}_{DE}=
\tilde{L}_{D}^{\r}\tilde{R}_{\r}^{B}f_{BC}^{A}\tilde{R}_{\l}^{C}\tilde{L}_{E}^{\l}
\end{equation}
Here $\tilde{E}_{BC}$ is the components of the inverse matrix of $\tilde{E}$.
In the case of non-abelian duality, the dual vector fields are trivial
(i.e. $\tilde{R}_{\r}^{A}\sim\delta_{\r}^{A}$ and
$\tilde{L}_{\r}^{A}\sim\delta_{\r}^{A}$ etc). Then
\begin{equation}
\partial^{A}\tilde{E}_{BC}=f_{BC}^{A}
\end{equation}
 For this case, the solution
is
\begin{equation}
\tilde{E}_{BC}={E}_{BC}^{0}+f_{BC}^{A}x_{A}
\end{equation}
where  we included spectator fields. ($A$ does not run through
spectator degrees of freedom and ${E}_{BC}^{0}$ depends only on
these spectators.)

\subsection{The B-field Gauge Symmetry}

In this section we briefly touch on the gauge symmetryfor the $B$
field  $\de B_{\m\n}=\d_{[\m}\L_{\n]}$ previously mentioned in
Sec. 2. The argument is applicable to $N=0,1,2$

In abelian T-duality this symmetry may be treated as an
enlargement of the duality group, at least in certain cases
\cite{Giveon:1988tt}. It would be interesting to understand if a
similar interpretation is also possible for Poisson-Lie duality.
We thus ask if this gauge symmetry is compatible with the
 condition \re{PLcnd}. For the $B$-field this condition
reads (in form language)
 \beq {\cal L}_{A}B={ i}_AH+d({
i}_AB)=\half\tilde{f}^{BC}_{A}\left({ i}_BB\wedge {
i}_CB+R_B\wedge R_C\right)~, \eeq{PLB} where $R_A\equiv
R^\n_AG_{\m\n}dX^\n$, and $i_A$ represents the contraction with
$R^\m_A$. Since the field strength $H$ is invariant, the variation
$\de B=d\L$ gives \beq d({
i}_Ad\L)+\tilde{f}^{BC}_{A}{i}_CB\wedge{ i}_Bd\L=0~. \eeq{PLBvar}

We first consider the possibility that the relation \re{PLBvar} is
in fact an identity. This can be shown to be the case if
$2\tilde{f}^{BC}_{A}{i}_CB=-\half f_{AC}^{~~B}R^C$, an equation
which may be solved to express $B_{AB}$ as a function of the
structure constants $f_{AC}^{~~B}$ and $\tilde {f}^{BC}_{A}$.
Clearly this represents a very special configuration.  Otherwise,
when \re{PLBvar} is not an identity, it may be interpreted as a
structure equation. Viewing it like this, it is immediately clear
that it generically gives
$\omega_A^{~B}=\tilde{f}^{BC}_{A}{i}_CB$ as a function of the
gauge parameter. This cannot be the case and we conclude that
there is an incompatibility. Perhaps it is possible to amend the
Poisson-Lie condition with terms that take care of this, but we
will not pursue this topic further here.

\section{Supersymmetric Examples}

In this section we present two examples which illustrates some of
the previous discussion. Generally, there are several different
ways to decompose a Drinfeld double into bi-algebras and an
organizing principle is needed
\cite{Snobl:2002kq},\cite{Snobl:2002jq}. Typically, in an
application the choice will be dictated by additional
requirements, e.g., tracelessness of the structure constants,
imposed to preserve the conformal invariance of string theory
\cite{VonUnge:2002ss}. Further, while there is a full
classification of all six-dimensional Drinfeld doubles
\cite{Snobl:2002kq}, a similar classification for the eight
dimensional doubles is lacking. Since these are the smallest
doubles of interest for $N=2$, looking for such examples will be
somewhat hampered by this lack of classification.

We take our
starting point in the well known example of the  $N=4$ supersymmetrical WZW model on $SU(2)\times
U(1)$, \cite{Rocek:1991vk}. We want to find a $N=1$ supersymmetric model instead, based on a
Drinfeld double with this group as part of the double. We find the double via a slight
generalization of the $SU(2),E_3$ double of Sfetsos. ($E_3$ is described in more detail below).
It would also be interesting to extend this to $N=2$. We shall see that although we will find an
almost complex structure, it fails to satisfy the conditions needed for $N=2$.

A group element is
\ber
g=\frac{e^{i\theta}}{\sqrt{\phi\bar{\phi}+\lambda\bar{\lambda}}}
\left(\begin{array}{cc}
\lambda & \bar{\phi} \\
-\phi & \bar{\lambda}
\end{array}\right)
\eer{gel}
where $\theta\equiv -\frac{1}{2}\ln(\phi\bar{\phi}+\lambda\bar{\lambda})$.
The right and left invariant frames are found from \re{LeRi} with generators
$t_{A}=(\frac{\sigma_{0}}{2},\frac{\sigma_{a}}{2})$;
$\sigma_{0}=\left(\begin{array}{cc}
1 & 0 \\
0 & 1
\end{array}\right)$;
$Tr(t_{A}t_{B})=\frac{1}{2}\delta_{AB}$;
$\m=\{\phi,\bar{\phi},\lambda,\bar{\lambda}\}$; $A=\{0,a\}$;
$a=\{1,2,3\}$. These vector fields generate the $su(2)\oplus u(1)$
algebra 
\ber [L_{0},L_{b}]&=&0;\ \ \ \ \
[L_{a},L_{b}]=i\epsilon_{abc}L_{c} \\ \nonumber
~[R_{0},R_{b}]&=&0;\ \ \ \ \ [R_{a},R_{b}]=-i\epsilon_{abc}R_{c}
\\ \nonumber
 ~[L_{A},R_{B}]&=&0~, 
\eer{sualg} 
and their explicit form is
given in the appendix.

The algebra of the other component in the Drinfeld double ${\cal
D}^2$ is that of $e_3 \oplus u(1)$: \beq
[t_i,t_j]=[t_0,t_i]=0~,\qquad [t_3,t_j]=t_i~, \quad i=1,2~.
\eeq{e3alg} 
The structure constants $f_{AB}^{~~C}$ and $\tilde
f^{AB}_{~~C}$ may be read off from (6.48) (left frames) and
\re{e3alg}, respectively. Defining the generators of the ${\cal
D}^2$ algebra according to
 \ber T_a&=&\left({\frac {\s_a}
2},{\frac {\s_a} 2}\right)~, T_0=\left({\frac 1 2},{\frac 1
2}\right)\\ \nonumber
 \tilde T^1&=&\left(\s_+ ,-\s_-\right)~,
\tilde T^2=-i\left(\s_+ ,\s_-\right)\\ \nonumber
 \tilde T^3&=&\left({\frac
{\s_3} 2},-{\frac {\s_3} 2}\right)~, \tilde T^0=\left({\frac 1
2},-{\frac 1 2}\right)~, \eer{Dgens} where $\s_a~,a=1,2,3$ and
$\s_\pm$ refer to the usual Pauli matrices and their $\pm$
combinations. With the definitions in (6.50), the generators
satisfy conditions \re{algebra}. The invariant product $\langle~
|~\rangle$ needed on the double is defined as \beq \langle
(A,B)|(C,D)\rangle \equiv \langle A|B\rangle-\langle C|D\rangle~,
\eeq{inpro} where $(A,B)\in {\cal D}^2$ and $\langle A|B\rangle=2
tr AB$. With \re{inpro} the generators in  (6.50) satisfy the
condition \re{innerprod}.

Having found a double and the left and right frames on one of the
components, we plug the frames into \re{inversePL} and solve it.
The solution is given by \beq
E^{AB}=E^{AB}_0+\Pi^{AB}=\eta^{AB}+c^{AB}+\Pi^{AB}~, \eeq{Esln}
where $\eta^{AB}$ is a constant symmetric matrix, $c^{AB}$ is a
constant antisymmetric matrix and $\Pi$ an antisymmetric
coordinate-dependent matrix which reads \beq \Pi={\frac 1
{D^2}}\left(\begin{array}{cccc} 0&0&0&0\\
0&0&-2i\l\bar\l&-\bar{\l}\bar{\phi}+\l\phi\\
0&2i\l\bar\l&0&-i(\bar{\l}\bar{\phi}+\l\phi)\\
0&\bar{\l}\bar{\phi}-\l\phi&i(\bar{\l}\bar{\phi}+\l\phi)&0
\end{array}\right)~,
\eeq{pi}
where $D^2\equiv \l\bar\l+\phi\bar\phi$. The solution \re{Esln} is the inverse of the one given
for the general case in \re{KSsln1}.
To write down the sigma model, we need the inverse $E_{AB}$ of the solution in \re{Esln},
$E_{AB}E^{BC}=E^{CB}E_{BA}=\de_A^C$. Introducing $\th^{AB}\equiv c^{AB}+\Pi^{AB}$ and using the
notation for $E_{AB}=G_{AB}+B_{AB}$ introduced in Sec.1, we find the following relations useful
in looking for the inverse
\ber
&&\eta^{AB}B_{BC}+\theta^{AB}G_{BC}=0\cr
&&\eta^{AB}G_{BC}+\theta^{AB}B_{BC}=\delta^A_C
\eer{rlns}
This implies that\footnote{Since $B$ is antisymmetric, the symmetric part of the RHS of
(\ref{bres}) has to vanish, which one can check that it does, writing it in terms of $E$ and
$E^{-1}$.}

\beq B_{AB}=-\eta_{AC}\theta^{CD}G_{DB}, \eeq{bres} where
$\eta_{AB}$ is the (constant) inverse of $\eta^{AB}$, and that the
inverse of $G$ is \beq
G^{AB}=\eta^{AB}-\theta^{AC}\eta_{CD}\theta^{DB}. \eeq{gres}
Equivalently \beq
G_{AB}=(\eta-\theta\eta\theta)_{AB}^{-1}=(E^{-1}\eta(E^{-1})^T)_{AB}.
\eeq{sum}
Inserting \re{Esln} into these relations we calculate
$E^{AB}$ and hence find a $N=1$ sigma model and its dual on the
double by inserting the result into \re{Gaction}.  The various
$N=1$ supersymmetric models possible are determined by the choices
of $\eta^{AB}$ and $c^{AB}$ in \re{Esln}. We present the result
for two different choices.

To find the explicit form of the double is straightforward. When
we  know $E^{AB}$ in \re{Esln}, we compute $\a $ and $\beta$ in
(5.35) using (6.50) and the invariant product (6.51).
In doing this, we also need to coordinatize the dual group
elements $\tilde g$. Finally \re{KSsln2} yields the dual metric
and $B$-field.

The supersymmetric actions result from inserting $E$ or $\tilde E$
into \re{susyaction}.

\subsection{Example I}

If we choose $\eta^{AB}=\de^{AB}$ and $c^{A0}=0$, we find
\beq
G_{AB}=\Delta^{-1}\left(\begin{array}{cccc}
\Delta&0&0&0\\
0&D^4-s^2_+&-is_+s_-&-2ls_+\\
0&-is_+s_-&D^4+s^2_-&2ils_-\\
0&-2ls_+&2ils_-&D^4-4l^2
\end{array}
\right)~,
\eeq{ginv}
where we use the condensed notation
\ber
&&l\equiv -\lambda\bar\lambda\cr
&&s_\pm\equiv\bar\lambda\bar\phi\pm\lambda\phi)\cr
&&(D^4-4l^2-s_-^2+s_+^2)D^{-4}\equiv \Delta ~.
\eer{defs}
In the same notation, the antisymmetric tensor reads
\beq
B_{AB}\equiv\Delta^{-1}\left(\begin{array}{cccc}
0&0&0&0\\
0&0&-il&s_-\\
0&il&0&is_+\\
0&-s_-&-is_+&0
\end{array}
\right)~.
\eeq{rub}

\subsection{Example II}

In this example, with an eye towards $N=2$, we attempt to find a complex structure that  preserves the new metric.

The hermiticity condition
\re{biherm} will be satisfied for a metric $G_{\m\n}$ if the corresponding relation is satisfied for the Lie-algebra components $\j_A^{~B}G_{BC}\j_D^{~C}=G_{AD}$, and this is
 equivalent to preservation of the inverse \re{gres}. From \re{Esln} it may be shown that it is sufficient to require preservation of $E^{-1}$, (or equivalently of $E$), a relation that we discussed in the paragraphs surrounding \re{Ecndn}. In fact, choosing\footnote{One of several possibilities}
\beq
\j=\left(\begin{array}{cccc}
-i&-2iq-p(\phi/\bar\l-\bar\phi/\l)&0&0\\
0&i&0&0\\
0&0&-i&\phi/\bar\l-\bar\phi/\l\\
0&0&0&i
\end{array}
\right)~, \eeq{t2} gives a $\j$ which preserves $E^{-1}$ with
antisymmetric part $\th$, provided that $c^{02}=c^{03}=0$, and
symmetric part \beq \eta^{-1}=\left(\begin{array}{cccc} 0&0&0&n\\
0&0&p&q\\ 0&p&0&0\\ n&q&0&0
\end{array}
\right)~,
\eeq{eta2}
where we may take $n=1$ without loss of generality.
We thus have an almost complex structure associated with the sigma model given by this choice of parameters. Unfortunately
it does not pass the next test for $N=2$; it does not satisfy \re{NJcond}. In fact, a further check shows that it is not integrable, its Nijenhuis tensor \re{Nij} is non-zero. (In calculating these relations we need to go to the coordinate expressions.)
We finally record the expression for $E$ in this case (with $q=0=c^{01}$, $p=1$):

\beq
E=\left(\begin{array}{cccc}
2i((\bar\l\bar\phi)^2-(\l\phi)^2)/A_+A_-&-i(\bar\l\bar\phi+\l\phi)/A_+&(\bar\l\bar\phi-\l\phi)/A_-&1\\
-i(\bar\l\bar\phi+\l\phi)/A_-&0&-D^2/A_-&0\\
(\bar\l\bar\phi-\l\phi)/A_+&D^2/A_+&0&0\\
1&0&0&0
\end{array}
\right)~,
\eeq{E}
where
\beq
A_\pm\equiv -2i\l\bar\l\pm D^2~.
\eeq{defigen}
\bigskip

 {\bf Acknowledgement:} We are grateful to Fawad Hassan, Martin Ro\v cek, Rikard von Unge and
  Maxim Zabzine for useful
and stimulating discussions. U.L. gratefully acknowledges
participation in the programme "Non-commutative phenomena in
mathematics and theoretical physics" at SHS, Norwegian Academy of
Science and Letters 2002, and the hospitality of NTU Athens, where
part of this research was carried out. U.L. also acknowledges
support from VR-grant 650-1998368 and EU-grant HPNR-CT-2000-0122

\section{Appendix}
\subsection{The right- and left-invariant frames}

The components of the right-invariant forms are
\begin{eqnarray}
R^{0}_{\phi}&=&-\frac{i\bar{\phi}}{D^{2}};\ \ \ \ \
R^{0}_{\bar{\phi}}=-\frac{i\phi}{D^{2}};\ \ \ \ \
R^{0}_{\lambda}=-\frac{i\bar{\lambda}}{D^{2}};\ \ \ \ \
R^{0}_{\bar{\lambda}}=-\frac{i\lambda}{D^{2}} \nonumber \\
R^{1}_{\phi}&=&-\frac{\bar{\lambda}}{D^{2}};\ \ \ \ \
R^{1}_{\bar{\phi}}=\frac{\lambda}{D^{2}};\ \ \ \ \
R^{1}_{\lambda}=-\frac{\bar{\phi}}{D^{2}};\ \ \ \ \
R^{1}_{\bar{\lambda}}=\frac{\phi}{D^2} \nonumber \\
R^{2}_{\phi}&=&\frac{i\bar{\lambda}}{D^{2}};\ \ \ \ \
R^{2}_{\bar{\phi}}=\frac{i\lambda}{D^{2}};\ \ \ \ \
R^{2}_{\lambda}=-\frac{i\bar{\phi}}{D^{2}};\ \ \ \ \
R^{2}_{\bar{\lambda}}=-\frac{i\phi}{D^{2}} \nonumber \\
R^{3}_{\phi}&=&-\frac{\bar{\phi}}{D^{2}};\ \ \ \ \
R^{3}_{\bar{\phi}}=\frac{\phi}{D^{2}};\ \ \ \ \
R^{3}_{\lambda}=\frac{\bar{\lambda}}{D^{2}};\ \ \ \ \
R^{3}_{\bar{\lambda}}=-\frac{\lambda}{D^2}
\end{eqnarray}
and the components of the right-invariant vectors are
\begin{eqnarray}
R^{\phi}_{0}&=&\frac{i\phi}{2};\ \ \ \ \
R^{\phi}_{1}=-\frac{\lambda}{2};\ \ \ \ \
R^{\phi}_{2}=-\frac{i\lambda}{2};\ \ \ \ \
R^{\phi}_{3}=-\frac{\phi}{2} \nonumber \\
R^{\bar{\phi}}_{0}&=&\frac{i\bar{\phi}}{2};\ \ \ \ \
R^{\bar{\phi}}_{1}=\frac{\bar{\lambda}}{2};\ \ \ \ \
R^{\bar{\phi}}_{2}=-\frac{i\bar{\lambda}}{2};\ \ \ \ \
R^{\bar{\phi}}_{3}=\frac{\bar{\phi}}{2} \nonumber \\
R^{\lambda}_{0}&=&\frac{i\lambda}{2};\ \ \ \ \
R^{\lambda}_{1}=-\frac{\phi}{2};\ \ \ \ \
R^{\lambda}_{2}=\frac{i\phi}{2};\ \ \ \ \
R^{\lambda}_{3}=\frac{\lambda}{2} \nonumber \\
R^{\bar{\lambda}}_{0}&=&\frac{i\bar{\lambda}}{2};\ \ \ \ \
R^{\bar{\lambda}}_{1}=\frac{\bar{\phi}}{2};\ \ \ \ \
R^{\bar{\lambda}}_{2}=\frac{i\bar{\phi}}{2};\ \ \ \ \
R^{\bar{\lambda}}_{3}=-\frac{\bar{\lambda}}{2}
\end{eqnarray}
The components of the left-invariant forms are
\begin{eqnarray}
L^{0}_{\phi}&=&-\frac{i\bar{\phi}}{D^{2}};\ \ \ \ \
L^{0}_{\bar{\phi}}=-\frac{i\phi}{D^{2}};\ \ \ \ \
L^{0}_{\lambda}=-\frac{i\bar{\lambda}}{D^{2}};\ \ \ \ \
L^{0}_{\bar{\lambda}}=-\frac{i\lambda}{D^{2}} \nonumber \\
L^{1}_{\phi}&=&-\frac{\lambda}{D^{2}};\ \ \ \ \
L^{1}_{\bar{\phi}}=\frac{\bar{\lambda}}{D^{2}};\ \ \ \ \
L^{1}_{\lambda}=\frac{\phi}{D^{2}};\ \ \ \ \
L^{1}_{\bar{\lambda}}=-\frac{\bar{\phi}}{D^2} \nonumber \\
L^{2}_{\phi}&=&\frac{i\lambda}{D^{2}};\ \ \ \ \
L^{2}_{\bar{\phi}}=\frac{i\bar{\lambda}}{D^{2}};\ \ \ \ \
L^{2}_{\lambda}=-\frac{i\phi}{D^{2}};\ \ \ \ \
L^{2}_{\bar{\lambda}}=-\frac{i\bar{\phi}}{D^{2}} \nonumber \\
L^{3}_{\phi}&=&\frac{\bar{\phi}}{D^{2}};\ \ \ \ \
L^{3}_{\bar{\phi}}=-\frac{\phi}{D^{2}};\ \ \ \ \
L^{3}_{\lambda}=\frac{\bar{\lambda}}{D^{2}};\ \ \ \ \
L^{3}_{\bar{\lambda}}=-\frac{\lambda}{D^2}
\end{eqnarray}
and the components of the left invariant vectors are
\begin{eqnarray}
L^{\phi}_{0}&=&\frac{i\phi}{2};\ \ \ \ \
L^{\phi}_{1}=-\frac{\bar{\lambda}}{2};\ \ \ \ \
L^{\phi}_{2}=-\frac{i\bar{\lambda}}{2};\ \ \ \ \
L^{\phi}_{3}=\frac{\phi}{2} \nonumber \\
L^{\bar{\phi}}_{0}&=&\frac{i\bar{\phi}}{2};\ \ \ \ \
L^{\bar{\phi}}_{1}=\frac{\lambda}{2};\ \ \ \ \
L^{\bar{\phi}}_{2}=-\frac{i\lambda}{2};\ \ \ \ \
L^{\bar{\phi}}_{3}=-\frac{\bar{\phi}}{2} \nonumber \\
L^{\lambda}_{0}&=&\frac{i\lambda}{2};\ \ \ \ \
L^{\lambda}_{1}=\frac{\bar{\phi}}{2};\ \ \ \ \
L^{\lambda}_{2}=\frac{i\bar{\phi}}{2};\ \ \ \ \
L^{\lambda}_{3}=\frac{\lambda}{2} \nonumber \\
L^{\bar{\lambda}}_{0}&=&\frac{i\bar{\lambda}}{2};\ \ \ \ \
L^{\bar{\lambda}}_{1}=-\frac{\phi}{2};\ \ \ \ \
L^{\bar{\lambda}}_{2}=\frac{i\phi}{2};\ \ \ \ \
L^{\bar{\lambda}}_{3}=-\frac{\bar{\lambda}}{2}
\end{eqnarray}

    \end{document}